\documentclass{moriond}
\usepackage[T1]{fontenc}
\usepackage[utf8]{inputenc}
\usepackage{lmodern}
\usepackage{microtype}
\usepackage[usenames,dvipsnames]{xcolor}
\usepackage{amsmath,amssymb,amsfonts,mathtools,bm,mathbbol}
\usepackage[compat=1.1.0]{tikz-feynhand}
\def\be{\begin{equation}}
\def\ee{\end{equation}}
\def\bea{\begin{eqnarray}}
\def\eea{\end{eqnarray}}

\def\phaMinus{\phantom{-}}
\newcommand{\eq}[1]{\eqref{eq:#1}}
\newcommand{\fig}[1]{Fig.~\ref{fig:#1}}
\newcommand{\tab}[1]{Tab.~\ref{tab:#1}}
\newcommand{\tr}{\mathrm{tr}}

\newcommand{\MPl}{M_\text{Pl}}
\newcommand{\dt}{{\widetilde{\delta}}}

\newcommand{\BM}[1]{$\mathbf{BM_{#1}}$}
\newcommand{\Photo}{}

\begin{document}
\vspace*{4cm}
\title{SAFE $\mathbf{Z'}$ FOR LFUV}

\author{ TOM STEUDTNER }

\address{Fakultät Physik, TU Dortmund, Otto-Hahn-Str.~4, D-44221 Dortmund, Germany\\
Department of Physics and Astronomy, University of Sussex, Brighton, BN1 9QH, U.K.
}

\maketitle\abstracts{
It is demonstrated that accounting for lepton flavor universality violating anomalies in $B$ meson decays via $Z'$ models implies a Landau pole problem. A family of models is proposed that resolve both issues simultaneously, fix the metastability of the Higgs and are predictive up to the Planck scale.
\newline
\begin{flushleft}
       \scriptsize Preprint: DO-TH 22/10
\end{flushleft}
}

\section{Introduction}

A key prediction of the Standard Model of Particle Physics (SM) is that the distinctive flavor of different lepton generations originates only in their Yukawa interactions. Consequently, leptons are the same up to their masses and couplings to the Higgs.  
Lepton flavor universality can be probed with theoretical clean observables~\cite{Hiller:2003js}
\begin{equation}\label{eq:RKratiodef}
       R_{H} = \frac{\int_{q_\text{min}^2}^{q_\text{max}^2} \frac{\text{d}\mathcal{B}(B\rightarrow H \mu^+ \mu^-)}{\text{d}q^2} \text{d}q^2}{\int_{q_\text{min}^2}^{q_\text{max}^2} \frac{\text{d}\mathcal{B}(B\rightarrow H e^+ e^-)}{\text{d}q^2} \text{d}q^2}  \qquad \text{with} \qquad  H=K,\,K^*,\,\phi,\,X_s,\cdots,
   \end{equation}
featuring branching fractions of flavor changing neutral current (FCNC) transitions and the 
dilepton invariant mass squares $q^2$. The SM expectation puts these ratios $R_H$ close to unity as electrons and muons couple lepton universal. 
However, in recent years this paradigm has been challenged, most prominently by the measurements of
\begin{equation}
       R_K\left(1.1 \, \text{GeV}^2 < q^2 < 6.0 \,  \text{GeV}^2 \right)=0.846^{+0.044}_{-0.041}, 
\end{equation}
by the LHCb-collaboration~\cite{Aaij:2021vac}, which is at $3.1\,\sigma$ tension with the SM.
Further hints of new physics are also observed in $R_{K^*}$~\cite{Aaij:2017vbb} and many more branching ratios and angular observables~\cite{Aaij:2020nrf,Aaij:2020ruw} 
in $B\rightarrow K^{(*)} \mu^+ \mu^-$ processes, which are from here on referred to as $B$-anomalies. While there is a long and diverse list of models to account for these anomalies via a SM extension, this work focuses on a popular approach of heavy $Z'$ models. Such models feature a new $U(1)'$ gauge group with generation-dependent charges, giving rise to FCNCs after fermion mass rotation.
Here we address a critical issue that is often overlooked with these solutions -- the appearance of Landau poles below the Planck scale.

\paragraph{Landau Pole.} 
In order to explain the $B$-anomalies a tree-level coupling of $Z'$ to quarks and leptons is necessary. 
Therefore, $Z'$ can be produced and probed at colliders, unless the signature is suppressed by a heavy mass $M_{Z'}$. 
Current experimental data implies lower mass bounds, one obtains $M_{Z'} \gtrsim 5$~TeV for non-vanishing couplings to the first-generation quarks \cite{Sirunyan:2021khd}.
However, $b \to s\ell\ell$ amplitudes mediated by $Z'$ are always $\propto g_4^2/M_{Z'}^2$, where $g_4$ is the $U(1)'$ gauge coupling. If $M_{Z'}$ is heavy, $g_4$ has to be sufficiently large to account for the $B$-anomalies.
This nudges the $g_4$ towards the non-perturbativity limit, which is manifested by a Landau pole at scale $\mu_\text{Landau}$.
An upper bound for $\mu_\text{Landau}$ can be estimated in a model with minimal $U(1)'$ charges that still account for the anomalies. In this case, at least the left-handed muon (charge $F_{L_2}$) and left-handed $b$ quark $(F_{Q_3})$ as well as additional fermions to cancel gauge anomalies, have to couple to $Z'$. 
The relevant $b \to s\mu\mu$ transition
\begin{equation}\label{eq:40TeV}
 \begin{tikzpicture}[baseline={([yshift=-.5ex]current bounding box.center)}]
       \begin{feynhand}
              \vertex [particle] (b) at (0em,+1.5em) {$b$};
              \vertex [particle] (s) at (0em,-1.5em) {$s$};
              \vertex [particle] (z) at (3.5em,-0.75em) {$Z'$};
              \vertex [dot] (v1) at (2em,0em) {};
              \vertex [dot] (v2) at (5em,0em) {};
              \vertex [particle] (l1) at (7em,+1.5em) {$\mu$};
              \vertex [particle] (l2) at (7em,-1.5em) {$\mu$};
              \propag [fermion] (b)  to (v1);
              \propag [fermion] (v1) to (s);
              \propag [fermion] (l1) to (v2);
              \propag [fermion] (v2) to (l2);
              \propag [photon] (v1) to (v2);
       \end{feynhand}
 \end{tikzpicture} \quad\sim \quad  
       \frac{1}{\Lambda^2} \simeq V_{ts}^\ast V_{tb}^{\phantom\ast} \,F_{L_2} \, F_{Q_3} \frac{g_4(M_{Z'})^2}{M_{Z'}^2}\,
\end{equation}
then points towards a scale of new physics of roughly~\cite{Hiller:2021pul} $\Lambda \simeq 40$~TeV. 
The leading order renormalization group running $(4\pi)^{2} \beta_{g_4} = \tfrac12  B_4 \, g_4^3$ yields an estimate of the Landau pole
\begin{equation}\label{eq:rge-Landau}
       \ln \frac{\mu_\text{Landau}}{M_{Z'}} \simeq \frac{(4\pi)^2}{B_4\, g_4(M_{Z'})^2 }\,,
\end{equation}
where the one-loop coefficient $B_4$ generically increases with the $U(1)'$ charges and we have $B_4 \geq \tfrac{16}3\left(F_{L_2}^2 + 3\,F_{Q_3}^2\right) $ even in the minimal model described above. Combining this lower bound with \eq{40TeV} and \eq{rge-Landau} one obtains
\begin{equation}\label{eq:Landau}
       \ln \frac{\mu_\text{Landau}}{M_{Z'}} 
       \lesssim \frac{\sqrt{3} \pi^2}{2} \, V_{ts}^\ast V_{tb}^{\phantom\ast} \left(\frac{\Lambda}{M_{Z'}}\right)^2 \quad \Rightarrow \quad \mu_\text{Landau} \lesssim 10^{10}\,\text{TeV}\,,
\end{equation}
 which is well below the Planck scale $\mu_\text{Landau} \ll \MPl \approx 10^{16}$~TeV. However, as the pole moves logarithmically, it is many orders of magnitude lower in realistic theories.  This phenomenon occurs generically for all $Z'$ models. Moreover, as lower bounds for $M_{Z'}$ increase in the future, the situation becomes even direr.
For the time being, some models may still escape this problem, e.g. if $M_{Z'}$ is significantly lower than $5$~TeV as there is no $Z'$ coupling to light quarks, or for large mixing angles in the up- and down sector rotations, see~\cite{Bause:2021prv} for details. 

\paragraph{Planck Safety.} The Landau pole issue and the related loss of predictivity suggest that the majority of $Z'$ models are ruled out as proper UV completions. As an effective theory, it has a short range of validity and is bound to pass on the non-perturbativity problem to the next theory in its high-energy limit. 
A true UV completion would have to be theoretically consistent and free of poles or unphysical parameter regimes at least until the Planck scale, where quantum gravity enters the game. This is exactly what we will refer to as \textit{Planck safety}. Note that this paradigm also includes stability of the scalar potential. 
In the following, we construct $Z'$ models that explain the B-anomalies, move the Landau pole past the Planck scale, and stabilize the Higgs and BSM scalar potential altogether.

\section{Pole-free $\mathbf{Z'}$ Models}
\begin{table}[h]
       \centering
      \renewcommand*{\arraystretch}{1.0}
       \setlength\arrayrulewidth{1.3pt}
       \setlength{\tabcolsep}{1pt}
       \begin{tabular}{|lc|ccccc|}
         \hline
         Field & &  Gen. & $\, U(1)_Y\, $ & $\, SU(2)_L \, $ & $\, SU(3)_C\, $ & $\, U(1)'\, $ \\ \hline
         SM fermions & $Q_i$ & 3 & $+1/6$ & $2$ & $3$ & $F_{Q_i}$ \\ 
         & $L_i$ & $3$ & $-1/2$ & $2$ & $1$ & $F_{L_i}$ \\ 
         & $U_i$ & $3$ & $+2/3$ & $1$ & $3$ & $F_{U_i}$\\ 
         & $D_i$ & $3$ & $-1/3$ & $1$ & $3$ & $F_{D_i}$\\ 
         & $E_i$ & $3$ & $-1$ & $1$ & $1$ & $F_{E_i}$\\ 
         Higgs scalar & $H$ & $1$ & $+1/2$ & $2$ & $1$ & $0$ \\ \hline
         BSM fermions & $\nu$ & $3$ & $0$ & $1$ & $1$ & $F_{\nu_i}$\\
         & $\psi_L$ & $3$ & $0$ & $1$ & $1$ & $F_{\psi}$\\
         & $\psi_R$ & $3$ & $0$ & $1$ & $1$ & $F_{\psi}$\\
         BSM scalars & $S$ & $3\times 3$ & $0$ & $1$ & $1$ & $0$\\ 
         & $\phi$ & $1$ & $0$ & $1$ & $1$ & $F_\phi$\\ \hline
       \end{tabular}
       \caption{SM and BSM fields with multiplicities (number of generations) and  representations under  $U(1)_Y \times SU(2)_L \times SU(3)_C \times U(1)'$. }
       \label{tab:fields}
\end{table}
We consider $Z^\prime$ extensions of the  
SM with generation-dependent
$U(1)'$ charges $F_{X_i}$ for the SM quarks $(X=Q,\,U,\,D)$, leptons $(X=L,\,E)$ as well as three right-handed neutrino (RHN) fields $(X=\nu)$. 
The $U(1)'$ symmetry is broken spontaneously by a BSM scalar $\phi$, generating a heavy mass for the $Z'$ boson. 
To avoid interference of $U(1)'$ and electroweak symmetry breaking, $\phi$ is a SM singlet while the Higgs has no $U(1)'$ charge: $F_H = 0$.
Moreover, a $3 \times 3$ completely uncharged BSM scalar matrix $S$ and  $3$ generations of vector-like BSM fermions $\psi_{L,R}$ are included. The latter carry universal $U(1)'$ charges, but are chosen to be invisible under the SM gauge group.  
All matter fields and representations under the ${U(1)_Y  \times SU(2)_L \times SU(3)_C \times U(1)'}$ gauge group are given in \tab{fields}.
     The Yukawa sector 
     \begin{equation}\label{eq:Yuk}
     \begin{aligned}
         - \mathcal{L}_\text{yukawa} & =  Y^d_{ij}\, \overline{Q}_i  H D_j + Y^u_{ij}\, \overline{Q}_i  \widetilde{H} U_j + Y^e_{ij}\, \overline{L}_i  H E_j  + Y^\nu_{ij}\, \overline{L}_i  \widetilde{H} \nu_j  +  y\, \overline{\psi}_{L i} \,S_{ij} \, \psi_{Rj} + \text{h.c.}
     \end{aligned}
     \end{equation}
     consists of the SM interactions and the right-handed neutrino coupling $Y^\nu$ to the Higgs, as well as a pure BSM Yukawa vertex described by a single coupling $y$, protected by a $U(3)_{\psi_L} \!\times\! U(3)_{\psi_R}$ BSM flavor symmetry that is only softly broken\footnote{Further terms coupling $\phi$ with $\nu_i$ and generating a  Majorana mass after $U(1)'$ breaking may be possible but are excluded here by the choice of $U(1)'$ charges.}.
     The scalar potential consists of the quartic terms
     \begin{equation}\label{eq:V4}
       \begin{aligned}
           V^{(4)} &=  \lambda \,(H^\dagger H)^2 + s\,(\phi^\dagger \phi)^2  + u\,\tr(S^\dagger S S^\dagger S)   + v\, \tr(S^\dagger S) \tr(S^\dagger S) \\
           &\phantom{=} + \delta\,(H^\dagger H)\, \tr(S^\dagger S) + \dt\,(H^\dagger H)(\phi^\dagger \phi) + w\,(\phi^\dagger \phi)\, \tr(S^\dagger S)\,,
       \end{aligned}
       \end{equation}
     featuring scalar self-interactions (first line) and portal couplings (second line). Further massive terms are present, see \cite{Bause:2021prv} for details.
     
     \paragraph{Constraints.} There are many $U(1)'$ charges in our models, which are however constrained by theoretical requirements and experimental measurements. This includes the cancellations of gauge anomalies, which arise from $U(1)'^3$, $U(1)_Y \times U(1)'^2$, $U(1)_Y^2 \times U(1)'$, $SU(2)_L^2 \times U(1)'$, $SU(3)_C^2 \times U(1)'$  and $U(1)'$ -- gravity fermionic triangles. These six lengthy conditions are listed in \cite{Bause:2021prv}.
     Moreover, for each component of the SM and RHN Yukawa couplings to be non-zero, a condition on $U(1)'$ charges is implied. Overall, there would be too many constraints to fulfill for all couplings. The top Yukawa is most important due to its size, but we choose to account for at least all diagonal quark Yukawas, leading to the six conditions $0 = F_{Q_i} + F_H - F_{U_i} $ and $0 = F_{Q_i} - F_H - F_{D_i}$.
     Further, due to strict constraints from electroweak measurements~\cite{Ellis:2017nrp}, our $Z'$ does not couple to electrons $F_{L_1} =  F_{E_1} = 0$.
      To avoid constraints from $K^0$--$\bar{K}^0$--oscillations~\cite{Zyla:2020zbs}, off-diagonal $Z'$ couplings to $s$- and $d$-quarks are switched off via $F_{Q_1} = F_{Q_2}$  and $F_{D_1} = F_{D_2}$.
      There is a similar constraint from $B_s$-mixing which has been taken into account, along with an upper bound for the gauge-kinetic and scalar mixing parameters, see \cite{Bause:2021prv} for further details.

      \paragraph{Global Fit.} 

      The impact of our $Z'$ model below the electroweak scale is estimated using the weak effective field theory framework, where the following (axial)-vector dimension-six operators 
\begin{equation} \label{eq:Heff}
\begin{aligned}
\mathcal{H}_\text{eff}^{bs\mu\mu} \supset & -\frac{4\,G_F}{\sqrt{2}}\frac{\alpha_e}{4\pi} \Big[\overline{s}_{L} \left(V_{ts}^\ast V_{tb}^{\phantom\ast}\right) \gamma^\nu \,b_L \Big]\Big[ \overline{\mu}\, \gamma_\nu \left(c_9^{bs\mu\mu} + c_{10}^{bs\mu\mu} \gamma_5\right) \mu\Big] \\
& -\frac{4\,G_F}{\sqrt{2}}\frac{\alpha_e}{4\pi} \Big[\overline{s}_{R} \left(V_{ts}^\ast V_{tb}^{\phantom\ast}\right) \gamma^\nu \,b_R  \Big]\Big[ \overline{\mu}\, \gamma_\nu \left(c_9^{bs\mu\mu\,\prime} + c_{10}^{bs\mu\mu\,\prime} \gamma_5\right) \mu\Big]\,,
\end{aligned}       
\end{equation}
relevant for  $b \to s\, \mu^+\, \mu^-$ transitions are induced from integrating out the $Z'$. Splitting the Wilson coefficients into a SM and new physics part $
       c_{9,10}^{bs\mu\mu \,(\prime)} = C_{9,10}^{\text{SM}\, \mu \,(\prime) } + C_{9,10}^{\mu\,(\prime)} $, 
the latter can be fitted to experimental data. 
\begin{table}[h]
       \centering
       \setlength\arrayrulewidth{1.3pt}
        \setlength{\tabcolsep}{3pt}
       \begin{tabular}{|c|c|c|c|c|c|c|}
       \hline
       Dim. & Fit for & $C_9^\mu$ & $C_{10}^\mu$ & $C_9^{\mu \,\prime}$ & $C_{10}^{\mu \,\prime}$ & $\text{Pull}_ \text{SM}$ \\ \hline
       1d & $C_9^\mu$ & $-0.83 \pm 0.14$ & -- & -- & -- & $6.0 \, \sigma$ \\ \hline 
       1d & $C_9^\mu=-C_{10}^\mu$ & $-0.41 \pm 0.07$ &  $-C_9^\mu$ & -- & -- & $6.0 \, \sigma$\\ \hline 
       2d & $C_{9,10}^\mu$ & $-0.71 \pm 0.17$ & $0.20 \pm 0.13$ & -- & -- & $5.9 \, \sigma$\\ \hline 
       4d & $C_{9,10}^{\mu \,(\prime)}$ & $-1.07 \pm 0.17$ & $0.18 \pm 0.15$ & $0.27 \pm 0.32$ & $-0.28 \pm 0.19$ & $6.5 \, \sigma$ \\ \hline
       \end{tabular}
       \caption{Best fit values for the Wilson coefficients $C_{9,10}^{\mu \,(\prime)}$ in different new physics scenarios and their respective pull from the SM hypothesis.
       } 
       \label{tab:FitValues}
\end{table}
 We have employed the global fit from~\cite{Bause:2021ply}, see there for a detailed discussion.
 Several fit scenarios are collected in \tab{FitValues}. It is apparent that only $C_9^\mu$ is vital to explain $B$-anomalies, while the other Wilson coefficients are consistent with zero.
 In this spirit, we focus on the new physics scenarios with $C_{10}^\mu = 0$, $C_9^\mu = - C_{10}^\mu$ as well as $C_{9}^\mu$ and $C_{10}^\mu$ matched independently, while switching off the couplings to right-handed quarks $C_{9,10}^{\mu \, \prime} = 0$.
 To that end, we make the following assumptions for the mass rotations of up/down type singlets $U_{u,d}$ and doublets $V_{u,d}$. Firstly, the CKM rotation of quarks is solely in the down-sector, while the up-type quarks are mass-diagonal: $V_d \approx V_\text{CKM}$ and $V_u,\,U_u \approx \mathbb{1}$. This disallows large mixing angles in the flavor rotations between the up- and down-type sectors. Moreover, we also assume that flavor-changing $Z'$ couplings to the right-handed quarks are negligible, which are generated by off-diagonal elements of $U_d$.
Hence, we obtain for the matching of the new-physics Wilson coefficients $C_{9,10}^\mu$ 
\begin{equation}\label{eq:C910}
       \begin{aligned}
              C_{9,10}^\mu &= - \frac{\pi}{\sqrt{2} G_F \,\alpha_e} \left(\frac{g_4}{M_{Z'}}\right)^2\left(F_{Q_3} - F_{Q_2}\right)\left(F_{E_2} \pm F_{L_2}\right) \,,
       \end{aligned}
\end{equation} 
which can be fixed according to the global fit scenarios in \tab{FitValues}, yielding constraints on  $U(1)'$ charges and gauge coupling $g_4$ at the matching scale.
In our analysis, we consider 4 benchmark models featuring the distinctive NP scenarios $C_{10}^\mu = 0$ with either no RHNs (\BM{1}) or lighter $Z'$ that does not couple to first and second-generation quarks (\BM{4}),  $C_{10}^\mu = - C_{9}^\mu$ (\BM{2}) or both Wilson coefficients being fitted independently (\BM{3}). We have listed the $U(1)'$ charge assignment of each benchmark model in \tab{Benchmarks}.

      \begin{table}[h]
       \setlength\arrayrulewidth{1.3pt}
        \def\arraystretch{1.25}
        \setlength{\tabcolsep}{1pt}
           \centering
              \begin{tabular}{|c|ccc|ccc|ccc|ccc|ccc|ccc|c|c|c|}
              \hline
              Model & & $F_{Q_i}$& & & $F_{U_i}$ & & & $F_{D_i}$&  & & $F_{L_i}$&  & & $F_{E_i}$ & & & $F_{\nu_i}$&  & $F_H$ & $F_\psi$ & $F_\phi$\\
              \hline
              \BM1 & $  \phaMinus \frac{1}{20}$ &   $\phaMinus \frac{1}{20}$ & $-\frac{1}{10} $ & $ \phaMinus \frac{1}{20}$ & $\phaMinus \frac{1}{20} $ & $-\frac{1}{10}$ & $\phaMinus \frac{1}{20}$ & $\phaMinus \frac{1}{20} $ & $-\frac{1}{10}  $ & $ \phaMinus 0 $ & $-\frac{9}{10}$ & $\phaMinus \frac{9}{10}$ &
              $ \phaMinus 0$ & $-\frac{9}{10}$ & $\phaMinus \frac{9}{10} $ & 
              $ \phaMinus 0$ & $\phaMinus 0$ & $\phaMinus 0$ & 
              $0$ & $1$ & $\frac{1}{5}$ \\
              \BM 2 & $ -\tfrac14 $ & $-\tfrac14 $ & $\phaMinus \tfrac16 $ &
              $  -\tfrac14 $ &  $-\tfrac14 $ & $\phaMinus \tfrac16 $ & 
              $  -\tfrac14 $ &  $-\tfrac14 $ & $\phaMinus \tfrac16 $ & 
              $ \phaMinus 0$ & $\phaMinus 1$ & $\phaMinus 0 $ & 
              $ \phaMinus 0$ & $\phaMinus 0$ & $\phaMinus 1 $ & 
              $ \phaMinus \frac{1}{12}$ & $-\frac{1}{12}$ & $\phaMinus 1 $ & 
              $0$ & $\frac{11}{12}$ & $\tfrac19$ \\
              \BM 3 &$ -\tfrac18$ & $-\tfrac18$ & $\phaMinus 0  $ & 
              $ -\tfrac18$ & $-\tfrac18$ & $\phaMinus 0 $ & 
              $ -\tfrac18$ & $-\tfrac18$ & $\phaMinus 0 $ & 
              $ \phaMinus 0 $ & $\phaMinus \tfrac12 $ & $\phaMinus \tfrac14 $ & 
              $ \phaMinus 0 $ & $\phaMinus \tfrac14 $ & $\phaMinus \tfrac12 $ & 
              $ \phaMinus 0 $ & $\phaMinus \tfrac14 $ & $\phaMinus \tfrac12 $ & 
              $0$ & $1$ & $\tfrac18$ \\	
              \BM 4 & $\phaMinus 0$ & $\phaMinus 0$ & $\phaMinus\frac{1}{9}$ & $\phaMinus 0$ & $\phaMinus 0$ & $\phaMinus\frac{1}{9}$ & $\phaMinus 0$ & $\phaMinus 0$ & $\phaMinus\frac{1}{9}$ & 
              $\phaMinus0$ & $\phaMinus\frac{1}{3}$ & $-\frac{2}{3}$ & 
              $\phaMinus0$ & $\phaMinus\frac{1}{3}$ & $-\frac{2}{3}$ & 
              $\phaMinus0$ & $\phaMinus\frac{1}{3}$ & $-\frac{2}{3}$ & 
              $0$ & $1$ & $\tfrac16$ \\		
              \hline
              \end{tabular}
              \caption{$U(1)^\prime$ charge assignments $F_X, \, X=Q_i, U_i, D_i, L_i, E_i, \psi, H, \phi$ in the four benchmark models.}
              \label{tab:Benchmarks}
       \end{table}

\section{Planck Safety}

In spite of their differences, \BM{1\dots4} share many characteristics regarding Planck safety. 
Explaining the $B$-anomalies typically requires $\alpha_4(\mu_0) \simeq \mathcal{O}(10^{-2})$ around the BSM scale $\mu_0 \simeq M_{Z'}$, which implies a $U(1)'$ Landau poles in the $10 \dots 10^{5}$~TeV range which is well below our bound \eq{Landau}.
In order to obtain a QFT that is well-defined and predictive up to quantum gravity effects, this pole has to be moved past the Planck scale. 
For the same reason, the scalar potential has to remain stable up to  $\MPl$.
Both stipulations can be fulfilled in tandem due to Yukawa and scalar portal couplings \eq{Yuk}, \eq{V4}.

The running of gauge couplings is slowed down significantly by the BSM Yukawa interactions $\alpha_y = y^2/(4\pi)^2$, requiring 
\begin{equation}\label{eq:boundYukawa}
F_\psi^2\,\alpha_{y}(\mu_0) \gtrsim \mathcal{O}(10^{-1})
\end{equation}
to make the $U(1)'$ Landau pole appear only after the Planck scale.
At the same time, the absolute values of all other $U(1)'$ charges (cf.~\tab{Benchmarks}) have to be sufficiently small relative to $|F_\psi|$ in order to help 
slow down the growth of $\alpha_4 = g_4^2/(4\pi)^2$. 
This is particularly critical for $F_{Q_i,U_i,D_i}$, as quarks contribute to the growth of $\alpha_4$ with their color and isospin multiplicities. 
In turn, $F_{E_i,L_i}$ have  to remain sizable enough to accommodate for $B$-anomalies and cancel gauge anomalies. 

\begin{figure}
       \centering
       \includegraphics[scale=.6]{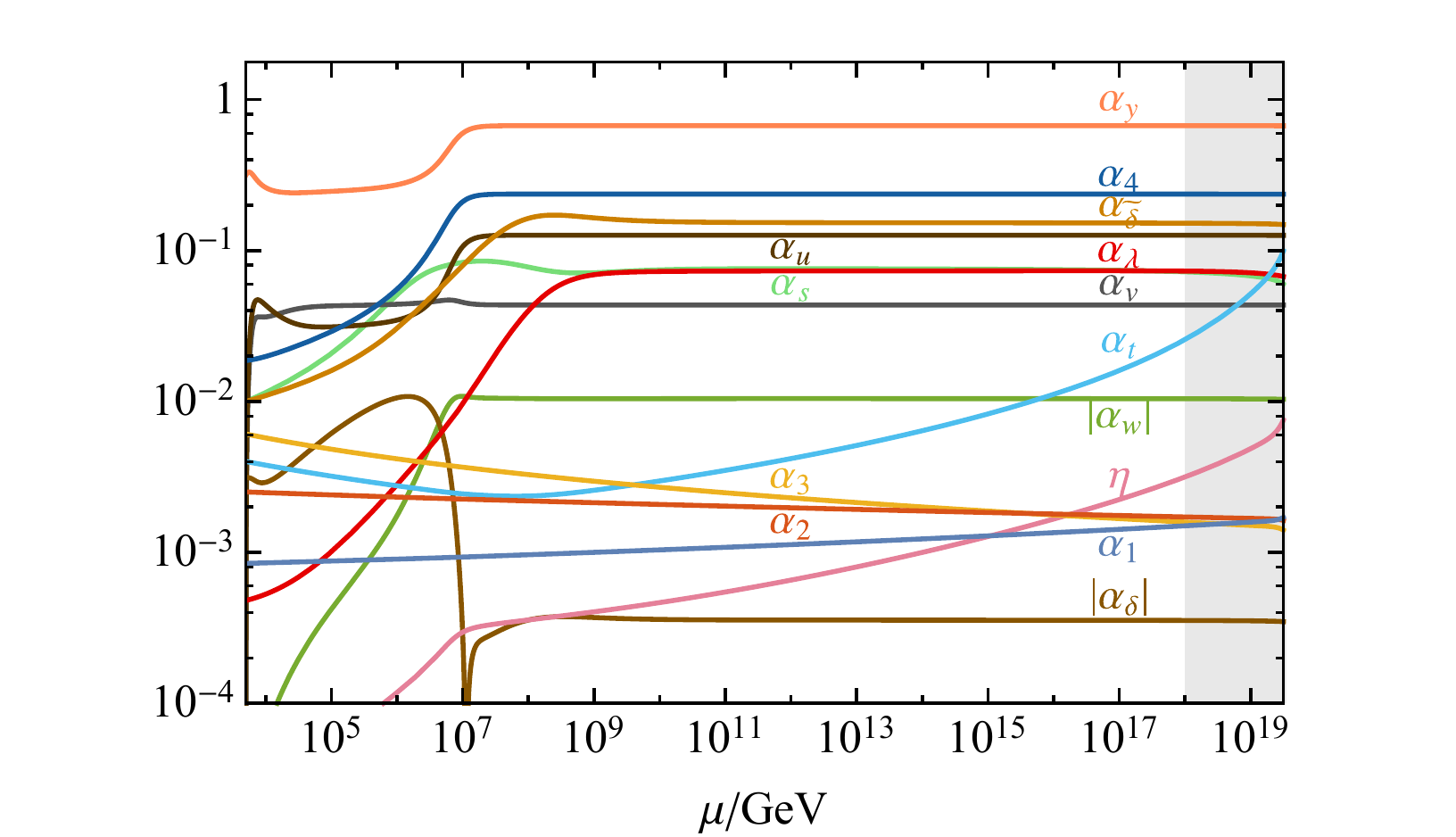}
       \caption{Example renormalization group evolution of gauge couplings $\alpha_i = g_i^2/(4\pi)^2$, Yukawa $\alpha_y = y^2/(4\pi)^2$ and quartic interactions $\alpha_u = u/(4\pi)^2$ as well as gauge-kinetic mixing parameter $\eta$ for \BM{1} from the initial scale $\mu_0=5$~TeV to the Planck regime (gray area).
       }
       \label{fig:run}
\end{figure}

On the other hand, the scalar potential -- including the metastable Higgs sector -- is stabilized through the portal couplings $\alpha_\delta = \delta /(4\pi)^2$ and $\alpha_\dt= \dt/(4\pi)^2$
that mediate the interactions between the SM Higgs as well as BSM scalars $S$ and $\phi$, see \eq{V4}. A rough estimate yields
\begin{equation}\label{eq:range}
10^{-3} \lesssim \alpha_{\delta}(\mu_0),\alpha_{\dt}(\mu_0) \lesssim 10^{-1}\,
\end{equation}
for either one or both of the couplings.
Interestingly, the stabilization of the potential and slowing of the $U(1)'$ gauge running are realized collaboratively.
For \BM{1}, this is depicted exemplarily in \fig{run}.
The sizable Yukawa and scalar portals often generate \textit{walking regimes}, where the evolution of couplings is dramatically slowed down due to the proximity of a pseudo fixed point. This locks the running of $U(1)'$ gauge and scalar self-interactions into a physical regime for many orders of magnitude. Other couplings, e.g. SM gauge interactions are not affected by the walking. The RG trajectory is bound to eventually escape this regime, which may happen before or after the Planck scale. At any rate, this phenomenon enables many such trajectories to pass through the $\MPl$ safely, without encountering instabilities or poles along the way.
\begin{figure}
       \centering
       \begin{tabular}{rl}
              \includegraphics[scale=.55]{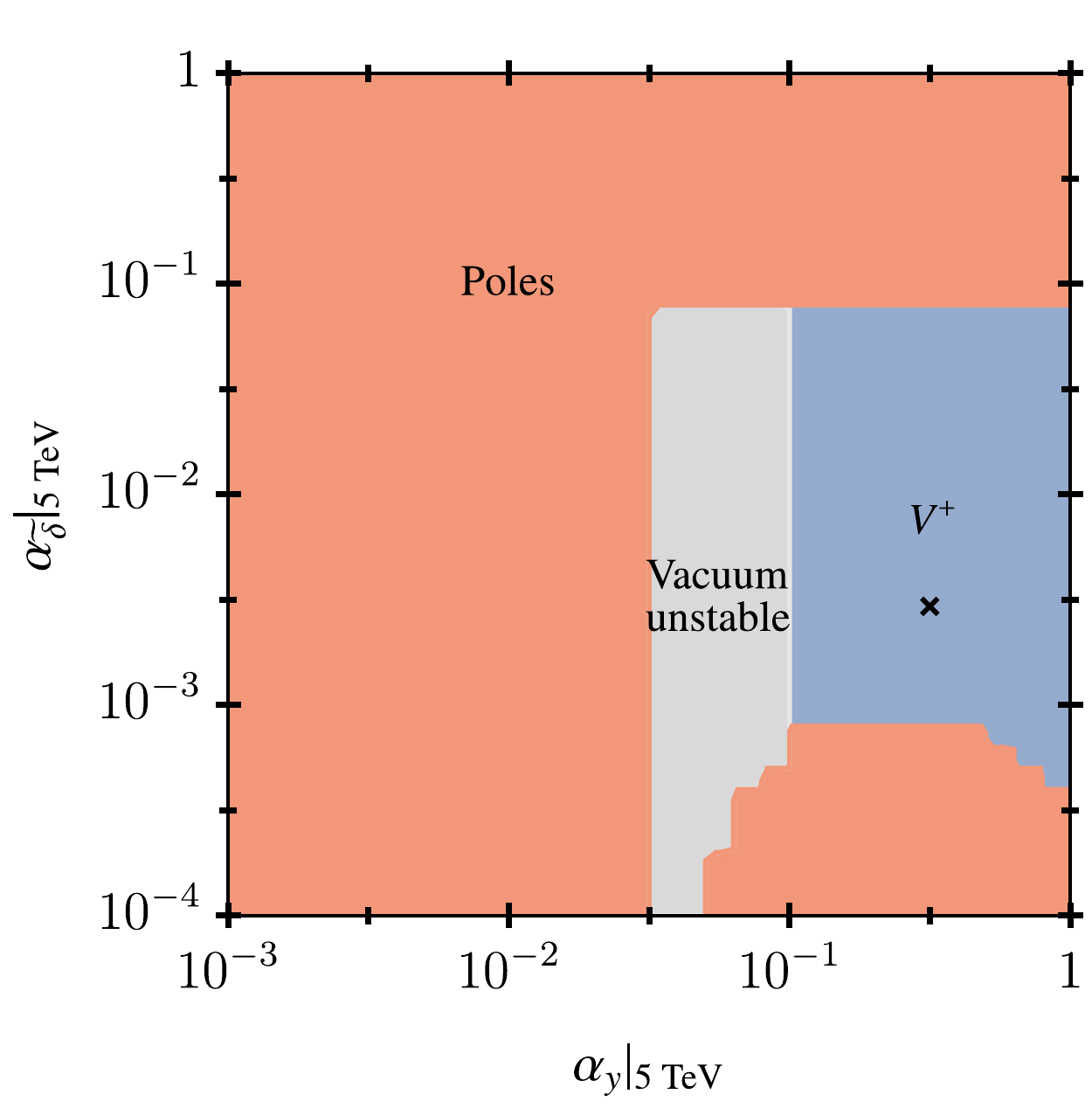} &
              \includegraphics[scale=.55]{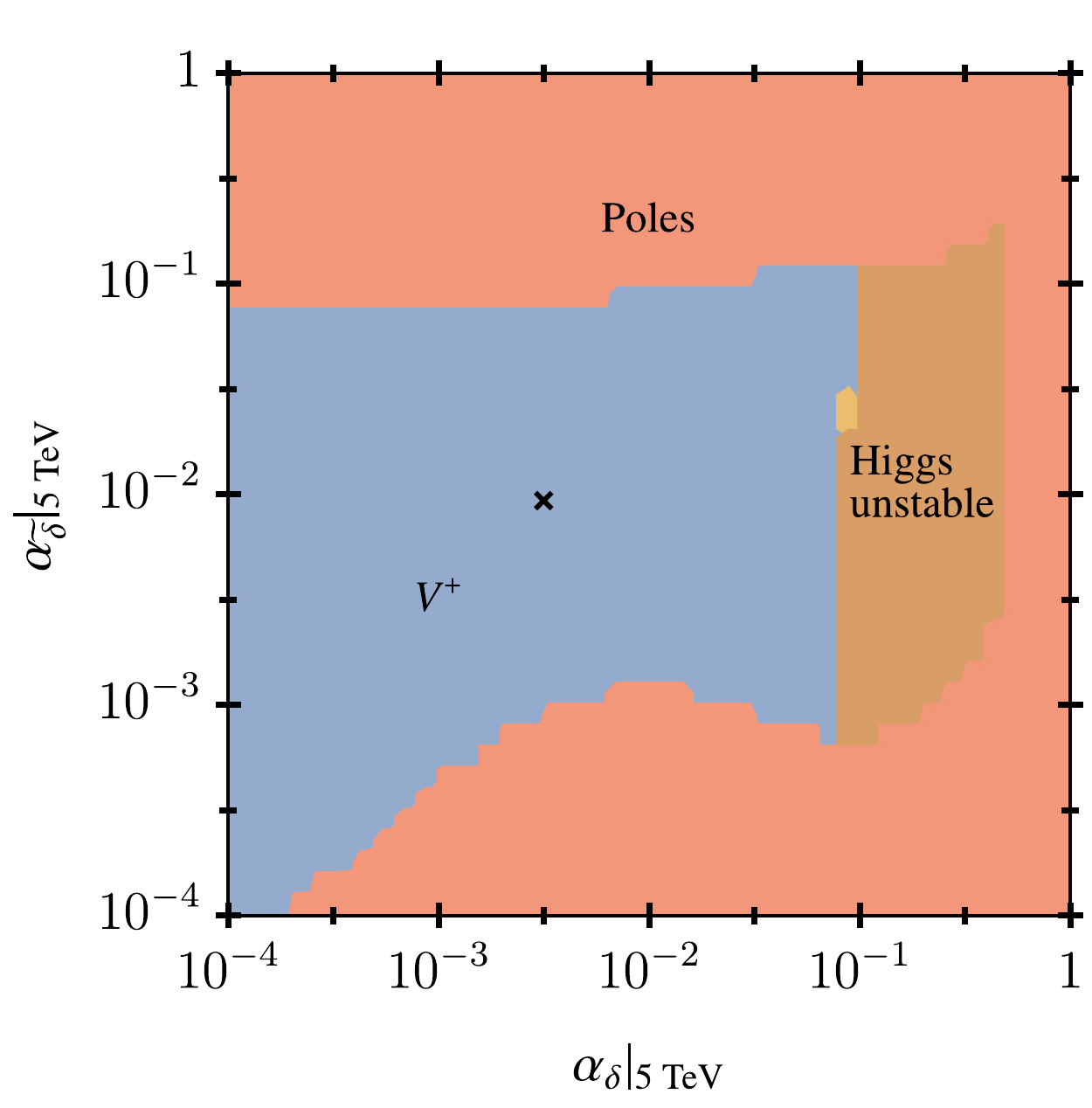}
       \end{tabular}
       \caption{Critical surface of parameters for \BM1 with matching scale $\mu_0 = 5$~TeV,
       projected onto the portal-Yukawa plane $\{\alpha_{y},\alpha_{\dt}\}|_{\mu_0}$  (left) and the portal-portal plane $\{\alpha_{\delta},\alpha_{\dt}\}|_{\mu_0}$ (right) while other BSM parameters are fixed to exemplary values. The black cross corresponds to the trajectory in \fig{run}. The color coding indicates if the vacuum at the Planck scale is unstable (gray)  or stable (blue), whether the Higgs is unstable (brown) or metastable ($-10^{-4}<\alpha_\lambda <0$, yellow), or whether poles have arisen (red).}
       \label{fig:surface}
\end{figure}
It is possible to determine the \textit{BSM critical surface}, i.e. the parameter ranges of BSM couplings at the matching scale that allow for Planck-safe RG trajectories. This is done exemplarily in \fig{surface}, where the blue area represents a slice of the BSM critical surface in accord with \eq{boundYukawa} and \eq{range}.

\section{Conclusion} 
In this work, we have demonstrated a systematic Landau pole problem with resolving $B$-anomalies via $Z'$ models. We have proposed a solution~\cite{Bause:2021prv} that accounts for many new-physics scenarios explaining the anomalies. These theories also well-defined and predictive until the Planck scale, and resolve the metastability of the SM Higgs.

\section*{Acknowledgments}
This contribution to the 2022 EW \& GUT session of the 56th Rencontres de Moriond is based on \cite{Bause:2021prv}. I would like to thank the organizers for their invitation as well as my coauthors of \cite{Bause:2021prv} for their collaboration.

\section*{References}  
\bibliography{ref-bib.bib}

\end{document}